\documentclass[12pt]{article}%
\usepackage{amsfonts}
\usepackage{cite}
\usepackage{amsmath}
\usepackage{amssymb}
\usepackage{bbm}
\usepackage{graphicx}
\usepackage[all,cmtip]{xy}
\usepackage{stackrel}%
\providecommand{\U}[1]{\protect\rule{.1in}{.1in}}
\makeatletter
\@addtoreset{equation}{section}
\makeatother

\begin{document}
\begin{titlepage}
\begin{center}
\renewcommand{\thefootnote}{\fnsymbol{footnote}}
{\Large{\bf Euclidean Supergravity in Five Dimensions}}
\vskip1cm
\vskip 1.3cm
Wafic A. Sabra and Owen Vaughan
\vskip 1cm
{\small{\it
Centre for Advanced Mathematical Sciences and Physics Department\\
American University of Beirut\\ Lebanon \\
ws00@aub.edu.lb,  owen.vaughan@physics.org}}
\vskip 1.3cm
\today
\end{center}
\bigskip
\begin{center}
{\bf Abstract}
\end{center}
We construct a 5D, ${\cal N} = 2$ Euclidean theory of supergravity coupled to vector multiplets.
Upon reducing this theory over a circle we recover the action of 4D, ${\cal N} = 2$ Euclidean supergravity coupled to vector multiplets.
\end{titlepage}\ \ \ \ \ \ \ \ \ \ \ \ \ \ \ \ \ \ \ \ \ \ \ \ \ \ \ \ \ \ \ \ \ \ \ \ \ \ \ \ \ \ \ \ \ 

\section{Introduction}

A comprehensive study of four-dimensional supersymmetric theories with
Euclidean spacetime signature has recently been conducted by Cort\'{e}s,
Mohaupt and collaborators in \cite{mohaupt1, mohaupt2, mohaupt3, mohaupt4},
where it is explained in detail how theories of $\mathcal{N}=2$
Euclidean\footnote{Following \cite{mohaupt1, mohaupt2, mohaupt3, mohaupt4} we
use the terminology $\mathcal{N}=2$ to refer to a supersymmetric theory with
eight real supercharges regardless of spacetime dimension or signature.}
vector multiplets (both rigid and local) are constructed. In the case of
Lorentzian signature it has been known for some time that the couplings of 4D,
$\mathcal{N}=2$ vector multiplets are restricted by supersymmetry such that
the scalar fields form a map into a target manifold with so-called special
geometry \cite{witvan84}. In the Euclidean case, it was found that the
analogues of special geometry are obtained, roughly speaking, by the
replacement of the complex fields of the theory by para-complex fields. See
\cite{mohaupt1} for further details on para-complex geometry.

The theory of 4D, $\mathcal{N}=2$ Euclidean vector multiplets coupled to
supergravity was constructed in \cite{mohaupt3} by the dimensional reduction
of 5D, $\mathcal{N}=2$ Lorentzian vector multiplets coupled to supergravity
\cite{GST} on a timelike circle. The couplings of
$\mathcal{N}=2$ Euclidean vector multiplets to supergravity are encoded in a
para-holomorphic prepotential $F$ that is homogeneous of degree two, and the
scalar fields form a map into a projective special para-K\"{a}hler target
manifold \cite{mohaupt3}. The Killing spinor equations were constructed in
\cite{ks}, again by dimensional reduction, and supersymmetric gravitational
instanton solutions were subsequently obtained in \cite{ginst}.

The construction of a theory of 5D, $\mathcal{N} = 2$ Euclidean vector
multiplets, either rigid or coupled to supergravity, is currently an open
problem. In this paper we propose a Lagrangian for an arbitrary number $n$ of
Euclidean vector multiplets coupled to supergravity. Aside from the spacetime
metric, the bosonic field content consists of $n$ real scalar fields $\phi
^{x}$ that parametrise a projective special real target manifold (this is the
same target geometry as in the Lorentzian case), and $(n + 1)$
abelian\footnote{In order to simplify the discussion we will only consider the
case of an abelian gauge group.} gauge fields $F^{i}$. In fact, the Lagrangian
looks almost identical to the Lorentzian case except for the fact that the
kinetic term for the gauge fields appears with the opposite sign to the
Lorentzian case.

The motivation for our choice of 5D, $\mathcal{N}=2$ Euclidean Lagrangian is
that upon dimensional reduction it produces the Lagrangian of 4D,
$\mathcal{N}=2$ Euclidean supergravity. We will see that it does not
immediately produce the Lagrangian presented in \cite{mohaupt3}, which was
obtained by reducing the 5D, $\mathcal{N}=2$ Lorentzian theory over time.
Instead, upon reducing our theory over a spacelike circle one produces a 4D
Lagrangian similar to \cite{mohaupt3} but with the `wrong' sign in front of
the gauge terms. However, we will demonstrate that in four dimensions with
Euclidean signature there is no preferred choice of sign in front of the gauge
fields, since theories with different signs in front of their gauge kinetic
terms can be mapped to one another by a duality transformation. This fact is
already known in the case of 4D Euclidean Einstein-Maxwell theory
\cite{instantons2, cosinst1, cosinst2}, and we will extend the argument to
include vector multiplets. This is already an interesting and useful result on
its own, and plays a key role in the construction of our 5D Euclidean theory.
The following schematic diagram summarises how the various theories in four
and five dimensions are related to one-another:
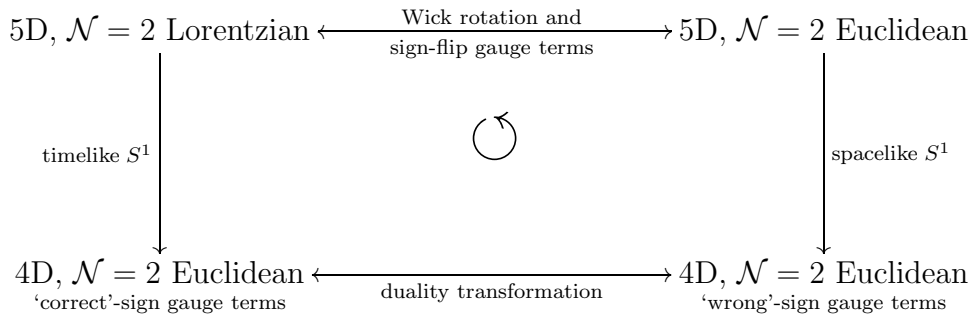
\vspace{1em}
\[
	\xymatrix
	{  
		\text{5D, ${\cal N} = 2$ Lorentzian} \ar[dd]_{\text{timelike} \; S^1} \ar@{<->}[rrrr]^{\text{Wick rotation and}}_{\text{sign-flip gauge terms}} & &  & &  \text{5D, ${\cal N} = 2$ Euclidean} \ar[dd]^{\text{spacelike} \; S^1}    \\	
		& &  \text{\huge $\circlearrowleft$} & &
		\\
		\stackrel[\text{`correct'-sign gauge terms}]{}{\text{4D, ${\cal N} = 2$ Euclidean}} \ar@{<->}[rrrr]_{\text{duality transformation}} & &  && \stackrel[\text{`wrong'-sign gauge terms}]{}{\text{4D, ${\cal N} = 2$ Euclidean}} 
	} 
\vspace{1em}
\]
Here the sign of the gauge terms refers to the overall sign in front of all
terms involving gauge field strengths in the Lagrangian.

Unlike in four dimensions, in five-dimensional Euclidean supergravity there is
only one allowed choice of sign in front of the gauge fields. The reason for
this is that after dimensional reduction the four-dimensional scalar fields
must take values in a para-complex target manifold (more specifically, a
projective special para-K\"{a}hler target manifold \cite{mohaupt1, mohaupt3}).
Recall that the four-dimensional scalar fields are constructed out of the five
dimensional scalar fields, the Kaluza-Klein scalar, and the components of the
gauge fields in the compact dimension. The only way one can obtain
para-complex scalar fields after reducing a 5D Euclidean theory over space is
to have a relative sign difference between the kinetic term for the scalar
fields and gauge fields in five dimensions. This is why in our 5D Euclidean
theory the terms involving gauge fields have a fixed overall sign, and that
this sign is opposite to the Lorentzian case.

The process in which we construct our five dimensional Euclidean theory by lifting from four dimensions is sometimes referred to as \emph{oxidisation},
since it is the reverse of the dimensional reduction procedure.
One may wonder why instead we did not try to obtain this theory from higher dimensions by dimensional reduction. 
For example, one could reduce an eleven-dimensional theory of Euclidean supergravity
 over a Calabi-Yau three-fold to five dimensions, in an analogous procedure
to the reduction of ten-dimensional Euclidean supergravity to four dimensions given in \cite{reduction}.
However, the minimal supersymmetry representation in eleven Euclidean dimensions has 64
real dimensions, and therefore reduction over a Calabi-Yau manifold (which breaks three-quarters of the supersymmetry) 
will result in a 5D Euclidean theory with a minimum of 16 real supercharges, which is not the theory 
that will be presented in this paper. 
It may be possible to obtain our theory by reducing eleven-dimensional theories with more exotic signatures, 
such as the $M^*$ and $M'$ theories developed by Hull in \cite{Hull:1998ym}, to five dimensions, however this will not be
investigated in this paper.

Our construction of a 4D, $\mathcal{N} = 2$ Euclidean Lagrangian with
the `wrong'-sign in front of the gauge fields also addresses a problem
encountered in \cite{mohaupt4} in the context of the $c$-map. In
\cite{mohaupt4} it was observed that there appeared to be a `fourth' $c$-map,
closely related to the Euclidean $c$-map, that was
mathematically well-defined but had no physical interpretation. Here we
provide such an interpretation: it is the map induced by the dimensional
 reduction of 4D, $\mathcal{N}= 2$ Euclidean supergravity with the
`wrong'-sign in front of the gauge fields over a spacelike circle. 
This will not be investigated any further in this paper.

This work is organised as follows: in section two we review
 five-dimensional Lorentzian supergravity and present our new
five-dimensional Euclidean supergravity theory. In section three we 
demonstrate that the sign in front of the gauge terms in the 4D, ${\cal N} = 2$
Euclidean vector multiplet Lagrangian can be positive or negative. The
theories with different signs are equivalent under a duality transformation.
In section four we show that the dimensional reduction of our five-dimensional Euclidean
theory to four dimensions results in a theory of 4D, ${\cal N}  = 2$ Euclidean supergravity
with the `wrong' sign in front of the gauge fields.

\section{Euclidean supergravity in five dimensions}

\label{sec:5DEuc}

\subsection{Review of Lorentzian theory}

Let us begin by reviewing the Lorentzian theory. The coupling of an arbitrary
number $n$ of vector multiplets to five-dimensional $\mathcal{N}=2$ supergravity
with Lorentzian spacetime signature was first considered in \cite{GST}. The
couplings in the bosonic fields are described by the so-called very special
geometry \cite{vp}, and are completely determined by a homogeneous cubic
polynomial $\mathcal{V}$ in $(n+1)$ real variables $h^{i}$, which is known as
the prepotential. The prepotential may be written as
\begin{equation}
\mathcal{V}={\frac{1}{6}}C_{ijk}h^{i}h^{j}h^{k}\,, 
\end{equation}
where $C_{ijk}$ are real constants symmetric in $i,j,k$. The physical theory
contains $n$ real scalar fields $\phi^{x}$ that parametrise the hypersurface
$\mathcal{V}=1$.

The Lagrangian is given by%
\begin{equation}
\hat{\mathbf{e}}^{-1}\hat{\mathcal{L}_{5}}=\frac{1}{2}\hat{R}-\frac{1}%
{2}G_{ij}\partial_{\hat{m}}h^{i}\partial^{\hat{m}}h^{j}+\frac{1}{4}G_{ij}%
F^{i}{}_{\hat{m}\hat{n}}F^{j}{}^{\hat{m}\hat{n}}+\frac{\hat{\mathbf{e}}^{-1}%
}{48}\,C_{ijk}\epsilon^{\hat{n}_{1}\hat{n}_{2}\hat{n}_{3}\hat{n}_{4}\hat
{n}_{5}}F^{i}{}_{\hat{n}_{1}\hat{n}_{2}}F^{j}{}_{\hat{n}_{3}\hat{n}_{4}}%
A^{k}{}_{\hat{n}_{5}},\ \label{eq:5DLagLor}%
\end{equation}
where $F^{i}{}_{\hat{m}\hat{n}}=2\partial_{\lbrack\hat{m}}A_{\hat{n}]}^{i}$
are the field-strength tensors. The moduli-dependent gauge coupling metric is
related to the prepotential by%
\begin{equation}
G_{ij}=-{\frac{1}{2}}\left(  {\frac{\partial}{\partial h^{i}}}{\frac{\partial
}{\partial h^{j}}}(\ln\mathcal{V})\right)  \Big|_{\mathcal{V}=1}={\frac{9}{2}%
}h_{i}h_{j}-{\frac{1}{2}}C_{ijk}h^{k}, \label{gc}%
\end{equation}
where the dual coordinates $h_{i}$ are defined by
\begin{equation}
h_{i}={\frac{1}{6}}C_{ijk}h^{j}h^{k}. \label{d}%
\end{equation}
In the Lorentzian theory the Killing spinor equations resulting from the
vanishing of the supersymmetry variation of the fermi fields in a bosonic
background are given by
\begin{align}
{{{\hat{D}}}}_{\hat{m}}{\hat{\varepsilon}}+{\frac{i}{8}}h_{i}\left(
\Gamma_{\hat{m}}{}^{{\hat{n}}_{1}{\hat{n}}_{2}}-4\delta_{{\hat{m}}}^{{\hat{n}%
}_{1}}\Gamma^{{\hat{n}}_{2}}\right)  F_{{\hat{n}}_{1}{\hat{n}}_{2}}^{i}%
{\hat{\varepsilon}}  &  =0,\nonumber\\
\left(  F^{i}-h^{i}h_{j}F^{j}\right)  _{{\hat{n}}_{1}{\hat{n}}_{2}}%
\Gamma^{{\hat{n}}_{1}{\hat{n}}_{2}}{\hat{\varepsilon}}-2i{\partial}_{\hat{m}%
}h^{i}\Gamma^{\hat{m}}{\hat{\varepsilon}}  &  =0. \label{eq:5DKSLor}%
\end{align}
Here ${{\hat{D}}}_{\hat{m}}=\partial_{\hat{m}}+\frac{1}{4}{\hat{\omega}%
}_{{\hat{m}},{\hat{n}}_{1}{\hat{n}}_{2}}\Gamma^{{\hat{n}}_{1}{\hat{n}}_{2}}$
is the five-dimensional covariant derivative.

Five-dimensional ${\cal N} = 2$ Lorentzian supergravity theories can be obtained by compactifying
eleven-dimensional supergravity over Calabi-Yau three-fold \cite{cad}. 
One may then interpret $\mathcal{V}$ as the intersection form of the Calabi-Yau
three-fold related to the overall volume of the Calabi-Yau three-fold.

\subsection{The 5D, $\mathcal{N} = 2$ Euclidean theory}

We propose that the theory of 5D, $\mathcal{N}=2$ Euclidean supergravity
coupled to $n$ vector multiplets is described by the Lagrangian%
\begin{equation}
\hat{\mathbf{e}}^{-1}\hat{\mathcal{L}_{5}}=\frac{1}{2}\hat{R}-\frac{1}%
{2}G_{ij}\partial_{\hat{m}}h^{i}\partial^{\hat{m}}h^{j}-\frac{1}{4}G_{ij}%
F^{i}{}_{\hat{m}\hat{n}}F^{j}{}^{\hat{m}\hat{n}}-\frac{\hat{\mathbf{e}}^{-1}%
}{48}\,C_{ijk}\epsilon^{\hat{n}_{1}\hat{n}_{2}\hat{n}_{3}\hat{n}_{4}\hat
{n}_{5}}F^{i}{}_{\hat{n}_{1}\hat{n}_{2}}F^{j}{}_{\hat{n}_{3}\hat{n}_{4}}%
A^{k}{}_{\hat{n}_{5}}\ .\label{et}%
\end{equation}
The coupling matrix $G_{ij}$ of the scalar fields and gauge fields remains the
same as in the Lorentzian case. The only difference between the Euclidean
Lagrangian \eqref{et} and the Lorentzian Lagrangian \eqref{eq:5DLagLor} is
that there is a relative sign difference in front of the third and fourth
terms, which involve the gauge field strengths $F^{i}$. We will see in section
\ref{sec:Reduction} that the choice of sign in front of the gauge kinetic
terms (the third term in \eqref{et}) is crucial in obtaining para-complex
scalar fields after dimensional reduction to four dimensions. Notice that the
sign in front of the Chern-Simons term (the fourth term) can be changed by
sending $A^{i}\rightarrow-A^{i}$, and is therefore not of relevance to the
discussion. We have selected the sign here for later convenience.

The Killing spinor equations of this Euclidean theory are given by
\begin{align}
{{{\hat{D}}}}_{\hat{m}}{\hat{\varepsilon}}-{\frac{1}{8}}h_{i}\left(
\Gamma_{\hat{m}}{}^{{\hat{n}}_{1}{\hat{n}}_{2}}-4\delta_{{\hat{m}}}^{{\hat{n}%
}_{1}}\Gamma^{{\hat{n}}_{2}}\right)  F_{{\hat{n}}_{1}{\hat{n}}_{2}}^{i}%
{\hat{\varepsilon}}  &  =0,\nonumber\\
\left(  F^{i}-h^{i}h_{j}F^{j}\right)  _{{\hat{n}}_{1}{\hat{n}}_{2}}%
\Gamma^{{\hat{n}}_{1}{\hat{n}}_{2}}{\hat{\varepsilon}}+2{\partial}_{\hat{m}%
}h^{i}\Gamma^{\hat{m}}{\hat{\varepsilon}}  &  =0. \label{kset2}%
\end{align}
Notice that these differ by a factor of $(-i)$ in front of all terms
containing gauge field strengths compared to the Lorentzian case
\eqref{eq:5DKSLor}. This compensates for the sign difference in the Lagrangian
between the two theories. We also have $\Gamma^{{\hat{n}}_{1}{\hat{n}}%
_{2}{\hat{n}}_{3}{\hat{n}}_{5}{\hat{n}}_{5}}=\epsilon^{{\hat{n}}_{1}{\hat{n}%
}_{2}{\hat{n}}_{3}{\hat{n}}_{5}{\hat{n}}_{5}}$, which is fixed by demanding
that the integrability conditions of the Killing spinor equations
(\ref{kset2}) are consistent with the field equations obtained from
({\ref{et}})

The motivation for our choice of the Lagrangian \eqref{et} and Killing spinor
equations \eqref{kset2} is that upon dimensional reduction over a spacelike
circle they correspond to the Lagrangian and Killing spinor equations of a
Euclidean theory of 4D, $\mathcal{N} = 2$ supergravity coupled to $n$ vector multiplets.
This will be demonstrated in section \ref{sec:Reduction}.

\section{Euclidean supergravity in four dimensions}

\label{sec:4DEuc}

In this section we will discuss Euclidean supergravity coupled to vector
multiplets in four spacetime dimensions. We will establish the following fact:
there is no preferred sign choice in front of the gauge terms in the 4D, ${\cal N} =2$
Euclidean vector multiplet Lagrangian. This will be important in section
\ref{sec:Reduction}, where we consider the reduction of five-dimensional
Euclidean supergravity to four dimensions. The key point is that in Euclidean
signature the equations of motion with either choice of sign can be mapped to
one-another by a duality transformation. This transformation consists of a
particular electric-magnetic duality transformation on the gauge fields (and
their duals), and a local field redefinition on the scalar fields.

In Euclidean Einstein-Maxwell theory it is known that one can have equivalent
formulation of the theory which differs by the sign of the gauge kinetic terms
\cite{instantons2, cosinst1, cosinst2}. It was argued in \cite{cosinst1} that
the reversal of sign in the Maxwell term in the action can be corrected at the
level of equations of motion by sending $\mathcal{F}$ to its dual. This
results in the stress-energy tensor of the Maxwell field changing sign. At the
level of the Killing spinor equations \cite{instantons2} the two formulations
are also related by sending $\mathcal{F}$ \ to its dual and using an
equivalent representation of Clifford algebra. In what follows, it will be
demonstrated that these arguments can be generalised to the four dimensional
Euclidean ${\cal N}=2$ theory with vector multiplets.

\subsection{Equations of motion}

It was shown in \cite{mohaupt3} that the theory of four-dimensional ${\cal N} = 2$
supergravity theory coupled to vector multiplets with Euclidean spacetime
signature may be described by the Lagrangian
\begin{equation}
\mathbf{e}^{-1}\mathcal{L}=\frac{1}{2}R-g_{ij}\partial_{\mu}z^{i}\partial
^{\mu}\bar{z}^{j}+\frac{1}{4}\left(  \mathrm{Im}\mathcal{N}_{IJ}%
\mathcal{F}^{I}\cdot\mathcal{F}^{J}+\mathrm{Re}\mathcal{N}_{IJ}\mathcal{F}%
^{I}\cdot\mathcal{\tilde{F}}^{J}\right)  , \label{action1}%
\end{equation}
where we used the notation $\mathcal{F}\cdot\mathcal{F}=\mathcal{F}%
_{ab}\mathcal{F}^{ab}$ etc. The Lagrangian has a similar form in the
Lorentzian case \cite{vanparis}, but the difference is that in the Euclidean
case the scalar fields $z^{i}$ are now para-complex and have a projective
special para-K\"{a}hler target manifold. It will be convenient to formulate
the Euclidean theory in terms of symplectic vectors $(L^{I},M_{I})$
that satisfy the symplectic constraint%
\begin{equation}
e\left(  \bar{L}^{I}M_{I}-L^{I}\bar{M}_{I}\right)  =1, \label{symp}%
\end{equation}
where $L^{I}=\mathrm{Re}L^{I}+e\mathrm{Im}L^{I}$ and $M_{I}%
= \frac{\partial F}{\partial L^I}.$ Here $\bar{e}=-e$ and $e^{2}=1$ and
\begin{equation}
\mathcal{N}_{IJ}:=\mathcal{R}_{IJ}+e\mathcal{I}_{IJ}:=\bar{F}_{IJ}%
-e\frac{N_{IK}L^{K}N_{JL}L^{L}}{L^{M}F_{MN}L^{N}}.
\end{equation}
Notice that $M_I = \mathcal{N}_{IJ}L^{J}$.
The constraint (\ref{symp}) is solved by setting%
\begin{equation}
L^{I}=e^{K(z,\bar{z})/2}X^{I}, \label{eq:XL}%
\end{equation}
where $K(z,\bar{z})$ is the K\"{a}hler potential of the theory. The geometry
of the the physical scalar fields $z^{i}$ of the vector multiplets is given by
a special K\"{a}hler manifold with K\"{a}hler metric
\begin{equation}
g_{i{\bar{j}}}=\frac{\partial^{2}K(z,\bar{z})}{\partial z^{i}\,\partial
{\bar{z}}^{j}}.
\end{equation}
Let us define
\begin{equation}
\mathcal{F}_{ab}^{\pm I}:=\frac{1}{2}\left(  \mathcal{F}_{ab}^{I}\pm
e\tilde{\mathcal{F}}_{ab}^{I}\right) ,\qquad\mathcal{G}_{Iab}%
^{+}:=\mathcal{N}_{IJ}\mathcal{F}_{ab}^{+J},\qquad\mathcal{G}_{Iab}%
^{-}:=\bar{\mathcal{N}}_{IJ}\mathcal{F}_{ab}^{-J}.
\end{equation}
The Maxwell fields stress-energy tensor appearing in the Einstein equations of
motion is given by%
\begin{equation}
T_{ab}=2\mathcal{I}_{IJ}\left(  \mathcal{F}_{ac}^{+I}\mathcal{F}%
_{\phantom{-J}b}^{-J\phantom{b}c}-\frac{1}{4}g_{ab}\mathcal{F}_{cd}%
^{+I}\mathcal{F}^{-Jcd}\right) , \label{ms}%
\end{equation}
and the
gauge field contribution to the scalar equations of motion is given by
the term%
\begin{equation}
e\left(  \partial_{I}\mathcal{N}_{JK}\mathcal{F}^{+J}\cdot\mathcal{F}%
^{+K}-\partial_{I}\bar{\mathcal{N}}_{JK}F^{-J}\cdot\mathcal{F}^{-K}\right).
	\label{eq:GaugeCont}
\end{equation}

Let us now make the field redefinition
\begin{equation}
\mathcal{F}_{ab}^{\prime+I}=\mathcal{G}{}_{Iab}^{+},\qquad\mathcal{F}%
_{ab}^{\prime-I}=\mathcal{G}_{Iab}^{-},\qquad L^{\prime I}=e\mathcal{N}%
_{IJ}L^{J},\text{ \ \ \ }M_{I}^{\prime}=eL^{I}. \label{fieldne}%
\end{equation}
The primed fields satisfy the same symplectic constraint
\[
1=e(\bar{L}^{I}M_{I}-L^{I}\bar{M}_{I})=e(\bar{L}^{\prime I}M_{I}^{\prime
}-L^{\prime I}\bar{M}_{I}^{\prime})\;.
\]
Notice that without gauge fields the scalar equations of motion and Einstein equations take the same form when written in terms of primed
fields. Under $GL(2n+2,\mathbb{R})$ transformations of the para-holomorphic
scalar fields
\begin{equation}
\left(
\begin{array}
[c]{c}%
L^{I}\\
M_{I}%
\end{array}
\right)  \mapsto\left(
\begin{array}
[c]{cc}%
U_{\;\;J}^{I} & Z^{IJ}\\
W_{IJ} & V_{I}^{\;J}%
\end{array}
\right)  \left(
\begin{array}
[c]{c}%
L^{J}\\
M_{J}%
\end{array}
\right)
\end{equation}
the gauge coupling matrix transforms as $\mathcal{N}\mapsto(V\mathcal{N}%
+W)(Z\mathcal{N}+U)^{-1}$ \cite{vanparis}$,$ and therefore $\mathcal{N}%
_{IJ}(L,\bar{L})=\mathcal{N}^{\prime IJ}(L^{\prime},\bar{L}^{\prime}).$ 
One may therefore interpret the \eqref{fieldne} as a strong-weak duality transformation of the gauge fields.

The stress-energy tensor of the Maxwell fields \eqref{ms} as well as the 
contribution of the gauge fields to the scalar equations of motion \eqref{eq:GaugeCont}
flip signs when expressed in terms of the primed fields. Moreover, the Bianchi
identities and Maxwell equations are interchanged under the field
redefinitions. We conclude that the equations of motion written 
in terms of the primed fields can be derived from a Lagrangian of the form \eqref{action1},
but with the opposite overall sign in front of all terms involving gauge fields
\begin{equation}
\mathbf{e}^{-1}\mathcal{L}=\frac{1}{2}R-g_{ij}\partial_{\mu}z^{i}\partial
^{\mu}\bar{z}^{j}-\frac{1}{4}\left(  \mathcal{I}_{IJ}\mathcal{F}^{I}%
\cdot\mathcal{F}^{J}+\mathcal{R}_{IJ}\mathcal{F}^{I}\cdot\mathcal{\tilde{F}%
}^{J}\right) \;.  \label{wa}%
\end{equation}

\subsection{Killing spinor equations}

The Killing spinor equations of 4D, $\mathcal{N}=2$ Euclidean supergravity, as
described by the Lagrangian (\ref{action1}), were constructed in \cite{ks}.
They are given by%
\begin{align}
\left(  {D}_{a}-{\frac{1}{2}}A_{a}\gamma^{5}\right)  \varepsilon+\frac{i}%
{4}\gamma.\mathcal{F}^{I}\left(  \mathrm{Im}L^{J}+\gamma^{5}\mathrm{Re}%
L^{J}\right)  \mathcal{I}_{IJ}\gamma_{a}\varepsilon &  =0,\nonumber\\
\frac{i}{2}\mathcal{I}_{IJ}\gamma.\mathcal{F}^{J}\left[  \text{Im}(g^{i\bar
{j}}\mathcal{D}_{\bar{j}}{\bar{L}}^{I})+\gamma_{5}\text{Re}(g^{i\bar{j}%
}\mathcal{D}_{\bar{j}}{\bar{L}}^{I})\right]  {\varepsilon}+\gamma^{a}%
\partial_{a}\left(  \text{Re}z^{i}-\gamma_{5}\text{Im}z^{i}\right)
{\varepsilon}  &  =0, \label{eq2}%
\end{align}
where%
\begin{equation}
\mathcal{D}_{^{\bar{j}}}{\bar{L}}^{I}=\left(  \mathcal{\partial}_{^{\bar{j}}%
}+\frac{1}{2}\mathcal{\partial}_{^{\bar{j}}}K\right)  {\bar{L}}^{I},
\end{equation}
and $A_{a}$ is the K\"{a}hler connection field and is expressed as
\begin{equation}
A_{a}=-\frac{e}{2}\left(\partial_{\alpha}K\partial_{a}z^{\alpha}-\partial
_{\bar{\alpha}}K\partial_{a}\bar{z}^{\bar{\alpha}}\right).
\end{equation}
It is convenient to express the above Killing spinor equations in terms of chiral
spinors. To do so we define
\begin{align}
\Gamma_{\pm}  &  =\frac{1}{2}\left(  1\pm e\gamma_{5}\right) , \nonumber
\end{align}
and then decompose $\varepsilon$ in terms of chiral and anti chiral parts
\begin{align}
\varepsilon=\varepsilon_{-}+\varepsilon_{+}, 
\qquad 
\text{where}
\qquad
\text{ \ \ }\Gamma_{\pm}\varepsilon_{\pm}  &  =\varepsilon_{\pm},\text{
\ \ }\Gamma_{\pm}\varepsilon_{\mp}=0. \nonumber
\end{align}
Note the identity
\begin{align}
\gamma_{5}\gamma
_{ab}=-{\frac{1}{2}}\epsilon_{ab}{}^{cd}\gamma_{cd}. \nonumber
\end{align}
With these conventions, we obtain from ({\ref{eq2}})\footnote{The two other
equations are obtained by sending $\varepsilon_{+}$ to $\varepsilon_{-}$ and
$e$ to $-e.$}%
\begin{align}
\left(  {D}_{a}-{\frac{e}{2}}A_{a}\right)  \varepsilon_{+}+{\frac{ie}{4}%
}\gamma.\mathcal{F}^{-I}L^{J}\mathcal{I}_{IJ}\gamma_{a}\varepsilon_{-}  &
=0,\label{q1}\\
{\frac{ie}{2}}\mathcal{I}_{IJ}\gamma.\mathcal{F}^{-J}g^{i{\bar{j}}%
}{\mathcal{D}}_{\bar{j}}{\bar{L}}^{I}\varepsilon_{+}+\gamma^{a}\partial
_{a}z^{i}\varepsilon_{-}  &  =0. \label{q2}%
\end{align}
Using the special geometry relations %
\begin{align}
\partial_{a}L^{I}-eA_{a}L^{I}  &  ={\mathcal{D}}_{i}L^{I}\partial_{a}%
z^{i},\nonumber\\
g^{\alpha\bar{\beta}}\mathcal{D}_{\alpha}L^{M}\mathcal{D}_{\bar{\beta}}\bar
{L}^{J}  &  =-\frac{1}{2}\mathcal{I}^{MJ}-\bar{L}^{M}L^{J},
\end{align}
we can write this as%
\begin{align}
\left(  D_{a}-\frac{1}{2}eA_{a}\right)  \varepsilon_{+}+\frac{i}{8}%
\gamma.(\,\mathcal{F}^{-J}M_{J}\mathcal{-G}_{J}^{-}L^{J})\gamma_{a}%
\varepsilon_{-}  &  =0, \\
-i\frac{1}{4}\gamma\cdot\left(  e\mathcal{F}^{-I}+\left(  \mathcal{F}%
^{-I}M_{I}-\mathcal{G}^{-I}L^{I}\right)  \bar{L}^{I}\right)  \varepsilon
_{+}+\gamma_{a}\left(  \partial_{a}L^{I}-eA_{a}L^{I}\right)  \varepsilon_{-}
&  =0. \label{sec}%
\end{align}

Let us now consider the field redefinition \eqref{fieldne}.
After making this field redefinition, defining\footnote{This is just a different representation of the Clifford algebra.} $\gamma_{a}=-i\gamma_{5}\gamma_{a}^{\prime
}$, and making use of the special geometry relation%
\[
dM_{I}-2\mathrm{Im}\mathcal{N}_{IJ}L^{J}A=\mathcal{\bar{N}}_{IJ}dL^{J},%
\]
we find (after dropping the primes) that the Killing spinor equations take the form%
\begin{align}
\left(  D_{a}-\frac{1}{2}eA_{a}\right)  \varepsilon_{+}-\frac{1}{8}%
\gamma.(\,\mathcal{F}^{-J}M_{J}\mathcal{-G}_{J}^{-}L^{J})\gamma_{a}%
\varepsilon_{-}  &  =0,\label{chiralone}\\
\frac{1}{4}\gamma\cdot\left(  e\mathcal{F}^{-I}+\left(  \mathcal{F}^{-I}%
M_{I}-\mathcal{G}^{-I}L^{I}\right)  \bar{L}^{I}\right)  \varepsilon_{+}%
+\gamma^{a}\left(  \partial_{a}-eA_{a}\right)  L^{I}\varepsilon_{-}  &  =0.
\label{chiraltwo}%
\end{align}
Notice that all terms involving gauge fields appear with an additional factor of $(-i)$.
These Killing spinor equations correspond to the Euclidean theory with the `wrong' sign of the gauge
terms, as described by the Lagrangian \eqref{wa}.

\section{Reduction from five to four dimensions}

\label{sec:Reduction}

In this section it will be demonstrated that the Lagrangian (\ref{wa}) (with cubic prepotential) and the
corresponding Killing spinor equations (\ref{chiralone}) and (\ref{chiraltwo})
can be obtained as a reduction of the five-dimensional Euclidean supergravity
theory (\ref{et}) and Killing spinor equations (\ref{kset2}) presented in
section \ref{sec:5DEuc}.

The reduction ansatz to four dimensions is given by
\begin{align}
\mathbf{\hat{e}}^{a}  &  =e^{-\phi/2}\mathbf{e}^{a},\text{ \ \ \ \ \ \ \ \ }%
\mathbf{\hat{e}}^{5}=e^{\phi}(dt-\sqrt{2}\mathcal{A}^{5}),\nonumber\\
A^{i}  &  =e^{-\phi}x^{i}\mathbf{\hat{e}}^{5}+\sqrt{2}\mathcal{A}^{i},\qquad
h^{i}\ =e^{-\phi}y^{i}.
\end{align}
All the fields are taken to be independent of the compact spatial dimension
labelled by index $5$, and the vector $\mathcal{A}^{5}$ has a vanishing
component along the fifth dimension. The non-vanishing components of the $d=5$ spin connection
 are given by
\begin{align}
{\hat{\omega}}_{5,5{\hat{a}}}  &  =e^{\frac{\phi}{2}}\partial_{a}\phi,\text{
\ \ \ \ \ \ \ \ \ \ \ \ \ \ }{\hat{\omega}}_{5,{\hat{a}}{\hat{b}}}%
={\frac{e^{2\phi}}{\sqrt{2}}}\mathcal{F}_{ab}^{5},\nonumber\\
{\hat{\omega}}_{{\hat{a}},5{\hat{b}}}  &  ={\frac{e^{2\phi}}{\sqrt{2}}%
}\mathcal{F}_{ab}^{5},\text{ \ \ \ \ \ }{\hat{\omega}}_{{\hat{a}},{\hat{b}%
}{\hat{c}}}=e^{\frac{\phi}{2}}\omega_{a,bc}+\frac{1}{2}e^{\frac{\phi}{2}%
}\left(  \eta_{ac}\partial_{b}\phi-\eta_{ab}\partial_{c}\phi\right)  .
\label{spin}%
\end{align}
where $\omega_{a,bc}$ are the $d=4$ spin connection associated with the basis
$\mathbf{e}^{a}$ and $\mathcal{F}^{5}=d\mathcal{A}^{5}$. The reduction of the
five-dimensional Hilbert-Einstein term in (\ref{et}) produces the following
term in the reduced Lagrangian
\begin{equation}
\sqrt{g}\left(  \frac{1}{2}R-\frac{3}{4}\partial_{a}\phi\partial^{a}\phi
-\frac{1}{8}e^{3\phi}\mathcal{F}^{5}.\mathcal{F}^{5}\right)  .
\end{equation}
Turning to the gauge fields, we have
\begin{equation}
F_{5{\hat{a}}}^{i}=-e^{-\phi/2}\partial_{{a}}x^{i},\text{ \ \ \ \ \ \ }%
F_{{\hat{a}}{\hat{b}}}^{i}=\sqrt{2}e^{\phi}(\mathcal{F}^{i}-x^{i}%
\mathcal{F}^{5})_{ab}, \label{mb}%
\end{equation}
and thus the kinetic terms of the gauge fields and the scalar fields in
(\ref{et}) reduce to%
\begin{align}
&  \sqrt{g}G_{ij}\left[  \frac{1}{2}e^{-2\phi}\partial_{{a}}x^{i}\partial
^{a}x^{j}+\frac{1}{2}e^{\phi}(\mathcal{F}^{i}-x^{i}\mathcal{F}^{5}%
)(\mathcal{F}^{j}-x^{j}\mathcal{F}^{5})\right] \nonumber\\
&  +\sqrt{g}\left(  -\frac{1}{2}e^{-2\phi}G_{ij}\partial_{a}y^{i}\partial
^{a}y^{j}+\frac{3}{4}\partial_{a}\phi\partial^{a}\phi\right)  .
\end{align}
The reduction of the Chern Simons term gives%
\[
S_{CS}=-\frac{1}{24}\int\sqrt{g}C_{ijk}\epsilon^{abcd}\left(  3x^{k}%
{}\mathcal{F}_{ab}^{i}\mathcal{F}^{j}{}_{cd}-3x^{i}x^{j}\mathcal{F}^{5}{}%
_{ab}\mathcal{F}^{k}{}_{cd}+x^{i}x^{j}x^{k}\mathcal{F}_{ab}^{5}\mathcal{F}%
_{cd}^{5}\right).
\]
Setting $G_{ij}\ =-2g_{ij}e^{2\phi}\,,$ and noting that $e^{3\phi}=\frac{1}%
{6}Cyyy,$  we obtain the reduced Lagrangian in four dimensions \ %
\begin{align}
\mathbf{e}^{-1}\mathcal{L}  &  =\frac{1}{2}R-g_{ij}\left(  \partial_{a}%
x^{i}\partial^{a}x^{j}-\partial_{a}y^{i}\partial^{a}y^{j}\right) \nonumber\\
&  -\frac{1}{6}Cyyy\left[  \frac{1}{4}\mathcal{F}^{5}\cdot\mathcal{F}%
^{5}+gxx\mathcal{F}^{5}\cdot\mathcal{F}^{5}+\,g_{ij}\,\mathcal{F}^{i}%
\cdot\mathcal{F}^{j}-2\,\left(  gx\right)  _{i}\!\mathcal{F}^{i}%
\cdot\mathcal{F}^{5}\right] \nonumber\\
&  -\frac{1}{12}\left[  3\left(  Cx\right)  _{ij}\mathcal{F}^{i}%
\cdot\mathcal{\tilde{F}}^{j}-3\left(  Cxx\right)  _{i}\mathcal{F}^{i}%
\cdot\mathcal{\tilde{F}}^{5}+\left(  Cxxx\right)  \mathcal{F}^{5}%
\cdot\mathcal{\tilde{F}}^{5}\right]  . \label{reduced}%
\end{align}
The dual field strength is $\mathcal{\tilde{F}}_{ab}=\frac{1}{2}%
\epsilon_{abcd}\mathcal{F}^{cd}$ and we have used the notation
\begin{equation}
Chhh=C_{ijk}h^{i}h^{j}h^{k},\text{ \ \ }\left(  Chh\right)  _{i}=C_{ijk}%
h^{i}h^{j},\text{ \ \ }\left(  Ch\right)  _{ij}=C_{ijk}h^{i}.
\end{equation}
The reduced Lagrangian is exactly what one obtains from the reduction of
Lorentzian five-dimensional supergravity on a time-like circle \cite{mohaupt3} but with
the sign of the gauge fields terms reversed. As in \cite{mohaupt3, ks}, it can
be demonstrated that the resulting theory describes the Lagrangian of 4D, $\mathcal{N}=2$ Euclidean 
supergravity, but with the opposite sign in front of the gauge terms. The prepotential is given by 
\begin{equation}
F=\frac{1}{6}C_{ijk}\frac{X^{i}X^{j}X^{k}}{X^{0}}\ , \label{pr}%
\end{equation}
and para-complex scalar fields by%
\begin{equation}
z^{i}=x^{i}-ey^{i}\;.%
\end{equation}

\subsection{Reduction of the Killing Spinor Equations}

We start with the two Killing spinor equations in (\ref{kset2}). The compact
component of the first equation gives, using our reduction ansatz%
\begin{equation}
\left[  e^{\frac{\phi}{2}}\partial_{a}\phi\gamma^{5}\gamma^{a}+{\frac{e^{\phi
}}{2\sqrt{2}}}\gamma.\left[  e^{\phi}\mathcal{F}^{5}-h_{i}\gamma^{5}%
{}(\mathcal{F}^{i}-x^{i}\mathcal{F}^{5})\right]  +h_{i}\gamma^{a}e^{-\phi
/2}\partial_{{a}}x^{i}\right]  {\hat{\varepsilon}}=0. \label{zc}%
\end{equation}

Using equation (\ref{zc}) and setting ${\hat{\varepsilon}=e}^{-\phi
/4}{\epsilon,}$ and after some Dirac matrices manipulations, we obtain
from the non-compact directions of the first equation%
\begin{equation}
D_{a}{\epsilon}+{\frac{3}{4}}h_{i}\gamma^{5}e^{-\phi}\partial_{{a}}%
x^{i}{\epsilon}-\frac{e^{\phi/2}}{8\sqrt{2}}\gamma.\left(  \mathcal{F}%
^{5}\gamma^{5}e^{\phi}+{3}h_{i}(\mathcal{F}^{i}-x^{i}\mathcal{F}^{5})\right)
\gamma_{a}\epsilon=0 \;. \label{kil41}%
\end{equation}
The second equation of (\ref{kset2}), using (\ref{zc}), gives%
\begin{equation}
\left[  \gamma^{a}\left(  \partial_{{a}}x^{i}+\gamma^{5}\partial_{a}%
y^{i}\right)  -\frac{e^{3\phi/2}}{2\sqrt{2}}\gamma.\left[  {h^{i}}e^{\phi
}\mathcal{F}^{5}-\gamma^{5}\left(  {3}h^{i}h_{j}-2\delta_{j}^{i}\right)
(\mathcal{F}^{j}-x^{j}\mathcal{F}^{5})\right]  \right]  \epsilon
=0.\label{kil42}%
\end{equation}
Using the relations between the four dimensional and five dimensional fields
together with the relations of special geometry \cite{ks}, it can be easily
shown that (\ref{kil41}) and (\ref{kil42}) \ are equivalent to
(\ref{chiralone}) and (\ref{chiraltwo}).

\bigskip

\textbf{Acknowledgements} : \ The work of W. S is supported in part by the
National Science Foundation under grant number PHY-1415659. We would like to
thank Jos\'{e} Figueroa-O'Farrill and Thomas Mohaupt for a useful discussion.


\begin{thebibliography}{99}                                                                                               %


\bibitem {mohaupt1}V. Cort\'es, C. Mayer and T. Mohaupt and F. Saueressig,
\textit{Special Geometry of Euclidean Supersymmetry I: Vector Multiplets,}
JHEP \textbf{03} (2004) 028.

\bibitem {mohaupt2}V. Cort\'es, C. Mayer and T. Mohaupt and F. Saueressig,
\textit{Special geometry of Euclidean supersymmetry II: hypermultiplets and
the c-map, }JHEP\textit{\ }\textbf{06}\textit{\ }(2005)\textit{\ }024.

\bibitem {mohaupt3}V. Cort\'es and T. Mohaupt, \textit{Special Geometry of
Euclidean Supersymmetry III: the local r-map, instantons and black holes},
JHEP \textbf{07} (2009) 066.

\bibitem {mohaupt4}V.~Cort\'{e}s, P.~Dempster, T.~Mohaupt and O.~Vaughan,
\textquotedblleft\textit{Special Geometry of Euclidean Supersymmetry IV: the
local c-map,} JHEP \textbf{10} (2015) 066.

\bibitem {witvan84}B. de~Wit and A. Van~Proeyen, \textit{Potentials and
Symmetries of General Gauged N=2 Supergravity: Yang-Mills Models}. Nucl. Phys.
\textbf{B245} (1984) 89.

\bibitem {GST}M. Gunaydin, G. Sierra and P. K. Townsend, \textit{The Geometry
of N=2 Maxwell-Einstein Supergravity And Jordan Algebras}, Nucl. Phys.
\textbf{B242} (1984) 244.

\bibitem {ks}J. B. Gutowski, and W.A. Sabra, \textit{Euclidean N=2
supergravity}, \ Phy. Lett. $\mathbf{B718}$ (2012) 610.

\bibitem {ginst}J. B. Gutowski and W. A. Sabra, \ \textit{Para-complex
geometry and gravitational instantons}, Class. Quantum Grav. \textbf{30}
(2013) 195001.

\bibitem {cosinst1}M. Dunajski, J. Gutowski, W. Sabra and P. Tod,
\textit{Cosmological Einstein-Maxwell Instantons and Euclidean Supersymmetry:
Anti-Self-Dual Solutions, }Class. Quant. Grav. \textbf{28} (2011) 025007.

\bibitem {instantons2}J. B. Gutowski and W. A. Sabra, \textit{Gravitational
Instantons and Euclidean Supersymmetry,} Phys. Lett. \textbf{B693} (2010) 498.

\bibitem {cosinst2}M. Dunajski, J. B. Gutowski, W. A. Sabra and P. Tod,
\textit{Cosmological Einstein-Maxwell Instantons and Euclidean Supersymmetry:
Beyond Self-Duality,} JHEP\textbf{\ 03} (2011) 131.

\bibitem {reduction}W. A. Sabra and Owen Vaughan, \textit{10D to 4D Euclidean
Supergravity over a Calabi-Yau three-fold}, Class. Quant. Grav. \textbf{33} (2015) 1033.

\bibitem{Hull:1998ym} C. M. Hull, \textit{Duality and the signature of space-time,}
  JHEP {\bf 9811} (1998) 017.  

\bibitem {vp}B. de Wit and A. van Proyen, \textit{Broken sigma-model
isometries in very special geometry}, Phys. Lett. \textbf{B293} (1992) 94.

\bibitem {cad}A. Ceresole, R. D'Auria, and S. Ferrara, \textit{11-Dimensional
Supergravity Compactified on Calabi-Yau Threefolds}, Phys. Lett. \textbf{B357}
(1995) 76.

\bibitem {vanparis}A.~Van Proeyen, $N=2$\textit{\ supergravity in }$d=4,5,6$
\textit{\ and its matter couplings}, extended version of lectures given during
the semester \textquotedblleft Supergravity, superstrings and
M-theory\textquotedblright\ at Institut Henri Poincar\'{e}, Paris, November
200 0; http://itf.fys.kuleuven.ac.be/$\sim$toine/home.htm\#B.
\end{thebibliography}
\end{document}